\documentstyle[12pt]{l-aa}
\input{psfig}
\begin{document}

\thesaurus{03(11.01.2 ; 13.07.2)}
\title{CGRO, Radio and Optical Observations of The Quasar NRAO~140}
\author{M. Maisack        \inst{1}
\and    R. Staubert       \inst{1}
\and    K. Otterbein      \inst{2}
\and    A. Witzel         \inst{2}
\and    S.J. Wagner       \inst{3}
\and    A. Heines         \inst{3}}
\offprints{M. Maisack}
\institute{Institut f\"ur Astronomie und Astrophysik der 
           Universit\"at T\"ubingen, Abteilung Astronomie, 
           72076 T\"ubingen, Germany
\and       Max Planck Institut f\"ur Radioastronomie, Auf dem H\"ugel 69, 
           D-53121 Bonn
\and       Landessternwarte K\"onigstuhl, D-69117 Heidelberg}
\date{Received 1995 ; accepted }
\maketitle

\begin{abstract}

We report on Compton Gamma Ray Observatory (CGRO), radio and optical 
observations of the radio-loud, superluminal quasar NRAO~140. The source is 
not detected (significance $3\sigma$) by any of the CGRO 
instruments OSSE, Comptel and EGRET. 
Radio observations simultaneous to the OSSE observation and optical 
monitoring over three-monthly intervals show no signs of extraordinary 
behaviour when compared to previous observations. We demonstrate that 
the CGRO non-detections do not require a spectral break at hard X-rays, 
but can be explained by a steep spectrum in the MeV range comparable to that 
of 3C~273.

\end{abstract}

\section{Introduction}

EGRET has detected $\approx$ 50 blazars at MeV and GeV energies 
(v.~Montigny et al. 1995a). It is generally believed that this emission 
originates in collimated outflows and is enhanced by relativistic beaming. 
Many of these objects show apparent superluminal motion. 
However, not all superluminal sources with flat radio spectra have been 
detected yet (v.~Montigny et al. 1995b). 
This may be due to either intrinsic physical reasons preventing the 
production of MeV $\gamma$-rays or their escape from the emission region, 
geometrical conditions such as misalignment or bending of the jet, or these 
sources may 
have been missed due to their variability, i.e. EGRET has only observed 
them in their low states. Since duty cycles of EGRET-detected blazars have 
a wide range (Heidt \& Wagner 1995, Wagner and Witzel 1995), 
this is a viable possibility to 
explain that many superluminal sources have not yet been detected at
$\gamma$-ray energies. 

To address this question, observations at energies of several tens to 
hundreds of keV may provide new insights, e.g. if a spectral break at these 
energies is detected. OSSE observations of this class of sources may 
provide insight into this issue. A number of blazars have already been 
detected by this instrument (McNaron-Brown et al. 1995). While most of the 
blazars observed so far by OSSE have been observed following their 
detection by EGRET, we follow a different approach and try to identify 
suitable candidates by their X-ray brightness. 

One of the most prominent of these unseen superluminal sources is the 
quasar NRAO~140 (Marscher and Broderick 1982), located at a redshift of 
z=1.258. This object is known for its 
brightness in X-rays (e.g. Marscher 1988, 
Ohashi et al. 1992) and its hard X-ray spectrum. Observations with EXOSAT 
(Marscher 1988), Ginga (Ohashi et al. 1992) and ASCA (Turner et al. 1995) 
have found power law spectra with photon indices of $\alpha$=1.6-1.7. No 
evidence for spectral 
hardening above $\approx$ 10~keV, indicative for the presence of a Compton 
reflected component, was found in the Ginga data. Extrapolating the EXOSAT, 
Ginga and ASCA spectra 
to the OSSE energy range ($>$ 50~keV), a positive detection was considered 
likely (see also Fig. 1). The observation of apparent superluminal motion 
made the object a good candidate for a Comptel or EGRET detection.

\begin{figure*}[thb]
 \psfig{figure=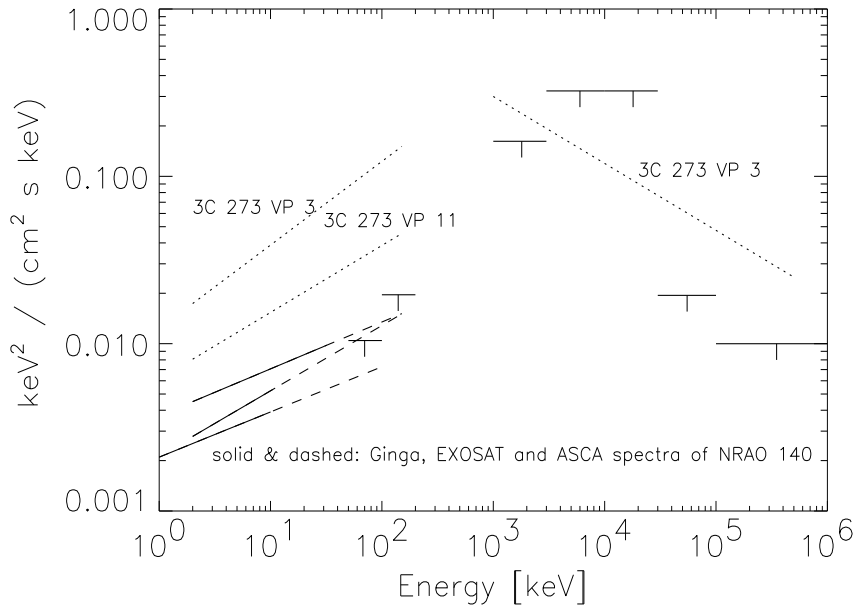,width=12cm,height=9.1cm}
 \caption[]{NRAO 140 Upper Limits vs. fluxes observed from 3C~273. 
Solid lines and dashed extensions show the EXOSAT and Ginga spectra 
of NRAO~140 and their extrapolations into the OSSE range. Dotted lines 
represent the  3C~273 spectra from VPs 3 and 8/11. Upper limits ($2\sigma$) represent 
NRAO~140 during the observations reported here.}
\end{figure*}

\section{Observations and Results}

\subsection{GRO observations}

CGRO consists of four instruments which 
span the $\gamma$-ray range from 20~keV to 20~GeV: BATSE, an uncollimated 
burst monitor 
(energy range $>$ 20~keV, Fishman et al. 1989); OSSE, a collimated phoswich 
type detector (50~keV - 10~MeV, Johnson et al. 1993), and the imaging 
instruments Comptel (0.75-30~MeV, Sch\"onfelder et al. 1993) and 
EGRET (50~MeV-30~GeV, Thompson et al. 1993). The latter three have been 
designed to observe point sources. 

In the course of this investigation, 
NRAO~140 was observed by OSSE during 1994 May 10-24 
(Viewing Period [VP] 326, 327), and by Comptel and EGRET during 
1994 Apr 26 - May~10 (VP 325). 
The source was not detected by any of these instruments. For the OSSE 
observation, the $2\sigma$ upper limit in the 50-150~keV 
range is 2.13$\times 10^{-3}$ photons / (cm$^2$ s MeV) for a net 
observation time of 465~ksec. This is just below 
the extrapolation of the Ginga spectrum, and compatible with the 
extrapolation of the EXOSAT and ASCA spectra. Contrary to non-detections of 
many 
Seyfert galaxies, the non-detection of NRAO~140 therefore does not require a 
spectral break at $\approx$ 100~keV. 
The EGRET upper limit from VP~325 (1.3$\times 10^{-7}$ photons / (cm$^2$ s) 
@ $>$ 100~MeV, 2$\sigma$) is comparable to that derived from several 
observations in Phase I (Fichtel et al 1994.) The Comptel upper limit 
could be improved by adding all available observations of NRAO~140 at various 
times.

\subsection{Radio and Optical Observations}

NRAO~140 was observed by the Effelsberg telescope during the OSSE 
observations in 1994 May 15-19 at 5~GHz. The observations at higher
frequencies -- namely 8.41~GHz -- were carried out within a period of 
10 days after the OSSE observations (Table 1). The measurements 
consisted of cross-scans (azimuth/elevation) through 
the source position. Corrections for gain and elevation effects were
applied using standard calibrator sources (e.\ g.\ NGC7027, 3C~286). The 
absolute calibration was performed according to Baars et al. (1987). The
flux densities were comparable to earlier
observations ($\approx$ 2~Jy, Table 1). An energy index
$\alpha^{5}_{8.4} = 0.2$ ($\rm S \sim \nu^{-\alpha}$) was 
calculated. Recent observations give a value of $\alpha^{5}_{8.4} 
= 0.3$ for 3C~273, similar to NRAO~140.

\renewcommand{\arraystretch}{1.25}
\begin{table}[thb]\centering
\caption{\label{Overview}Effelsberg Observations}
\begin{tabular}[c]{ccc}
\hline 
Date & Flux [Jy] & Error \\
\hline 
\multicolumn{3}{l}{5.0 GHz}\\
1994 May 15 & 1.96 & 0.03\\
1994 May 16 & 1.81 & 0.03\\
1994 May 17 & 1.94 & 0.03\\
1994 Jun 06 & 1.94 & 0.03\\
\hline
\multicolumn{3}{l}{6.0 GHz}\\
1994 May 19 & 1.78 & 0.03\\
\hline
\multicolumn{3}{l}{8.41 GHz}\\
1994 May 27 & 1.71 & 0.05\\
1994 May 28 & 1.74 & 0.05\\
1994 Sep 24 & 1.75 & 0.05\\
\hline
\end{tabular}
\end{table}

\begin{figure*}[thb]
 \psfig{figure=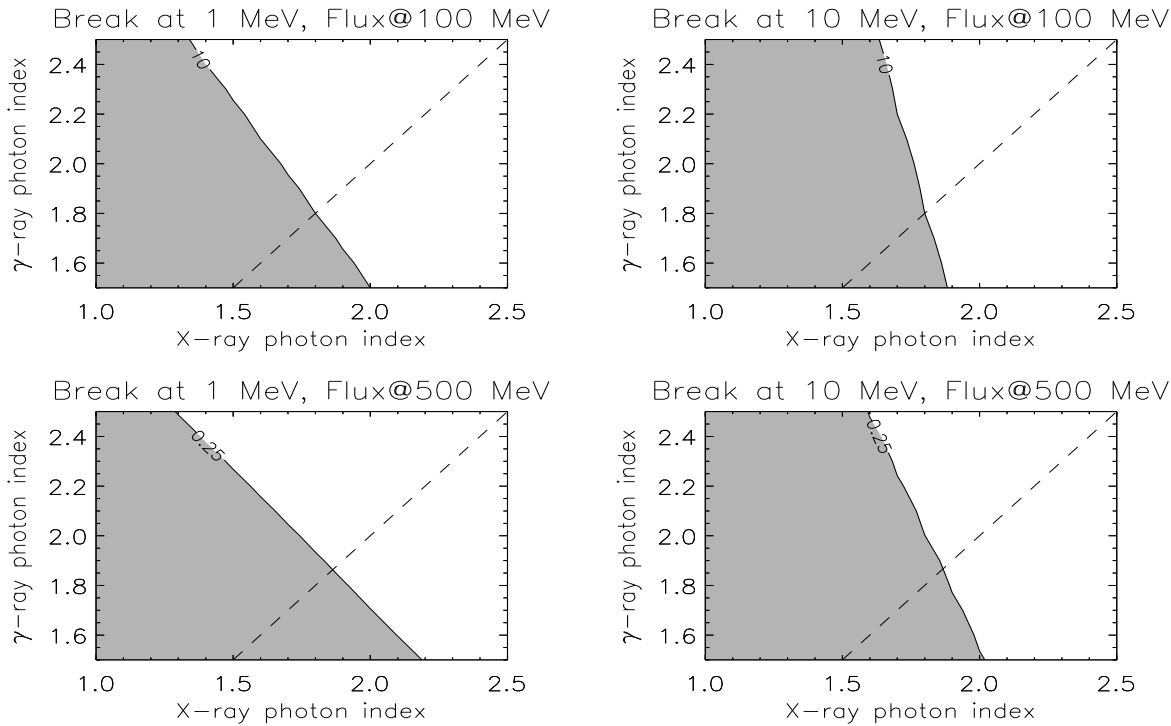,width=0.9\textwidth,height=9.9cm}
 \caption[]{Combinations of X-ray and $\gamma$-ray indices leading to an 
EGRET detection extrapolating typical ROSAT blazar fluxes 
($10^{-3} photons / (cm^2~s~keV)$ at 1~keV) to the EGRET 
energy range, assuming breaks at 1 or 10 MeV. Contours correspond to label 
values times $10^{-7} photons/ (cm^2~s~GeV)$, the 
detection sensitivities of EGRET at 100 and 500~MeV for a two-week 
observation 
. Dashed lines show 
$\alpha_x = \alpha_{\gamma}$. Combinations to the left of 
the contours (shaded areas) result in detection}
\end{figure*}

Relative photometry was performed at the Landessternwarte Heidelberg (see 
Table 2). A 
CCD attached to a 70~cm telescope was used for observations in the R-band 
(680nm). Individual epochs were tied together using non-variable field 
stars. In spite of the slight variations there is no indication of unusual 
activity. Both the average level as well as the scatter are typical for 
NRAO~140.

\renewcommand{\arraystretch}{1.25}
\begin{table}[thb]\centering
\caption{\label{Overview}Relative Photometry at LSW}
\begin{tabular}[c]{ccc}
\hline 
\multicolumn{3}{c}{Average: 17.42 $\pm$ 0.12}\\
\hline 
Date & R Mag. & Error\\
\hline
1994 Mar 03 & +0.13 & 0.03\\
1994 Mar 28 & -0.12 & 0.02\\
1994 Mar 29 & -0.11 & 0.02\\
1994 Jul 05 & -0.02 & 0.06\\
1994 Sep 19 & +0.12 & 0.02\\
\hline
\end{tabular}
\end{table}


\section{Discussion}

\subsection{NRAO 140}

In Fig. 1, we compare the CGRO upper limits of NRAO~140 to data of 3C~273 
(Johnson et al. 1995, v. Montigny et al. 1993). 
The dotted lines show the spectrum of 3C~273 during June 1991 (VP~3) 
measured by OSSE, Comptel and 
EGRET October 1991 (VP 8 and 11, OSSE; VP 11 Comptel and EGRET). 
During October 1991, 
3C~273 was about a factor of 3 less bright in the OSSE range than in VP 3.
At that time, it was not detected by Comptel and 
EGRET (Williams et al. 1995). The Comptel and EGRET upper limits for that 
observation of 3C~273 were comparable to the levels of NRAO~140. 
An X-ray/$\gamma$-ray spectrum similar to that of 3C~273 from VP 3, scaled 
by a factor of 5 or 10 to match the observed X-ray fluxes of NRAO~140, is 
compatible with the upper limits for NRAO~140 observed by CGRO. 
We conclude that for NRAO~140 a spectral break $\approx$ 100~keV 
is not required to explain the OSSE non-detection. Considering that OSSE 
has detected only about one third of the blazars that it has observed 
(McNaron-Brown et al. 1995), this indicates that OSSE detections even of 
bright blazars may only be feasible in a high state of emission. 
Furthermore, the Comptel 
and EGRET non-detections are also compatible with a 3C~273 type spectrum. 

\subsection{Other undetected superluminal sources}

v. Montigny et al. (1995b) have discussed the possible reasons for the 
non-detection of a number of prominent superluminal sources by EGRET. 
They consider 
variability and physical reasons such as misaligned jets or opacity. The 
proton initiated cascade model of Mannheim (1992) predicts that the proton 
to electron ratio governs the magnitude of gamma-ray emission which is low 
when the proton to electron ratio is close to or lower than 1. 
We adopt a phenomenological approach: 
the distribution of observed photon indices in the EGRET range shows that 
3C~273 has one of the steepest spectra ($\alpha_{\gamma}$=2.4) 
observed by EGRET so far, 
and, following the example of NRAO~140, we argue that sources with 
steep X-ray and $\gamma$-ray spectra and relatively low fluxes may 
fall below the EGRET sensitivity limit. 
To estimate whether this is a realistic assumption, we use known X-ray 
fluxes and broad-band high energy spectra of flat spectrum radio-loud quasars. 
Analysis of ROSAT archival data of blazars show that the observed 1~keV 
fluxes of these objects cluster around $10^{-3}$ photons/ (cm$^2$ s keV), 
with the exception of 3C~273, which is about a factor of 10 brighter. 
The distribution 
of ROSAT X-ray photon indices shows a scatter around a mean value 
of $\approx$ 2. 
We extrapolate power law spectra with indices similar to those observed by 
ROSAT from 1~keV to energies of 
1 or 10~MeV, where CGRO observations indicate that a break in the blazar 
spectra must be located (e.g. McNaron-Brown et al. 1995), and continue the 
extrapolation with a range of $\gamma$-ray indices broader than observed by 
EGRET to energies of several 100~MeV. 
We find that even if the X-ray spectrum is as hard as $\alpha_x$=1.5, 
a detection by EGRET becomes more and more unlikely as the gamma-ray photon 
index becomes steeper than $\alpha_{\gamma}$=2, unless the X-ray intensity 
increases by at least an order of magnitude, which is more than observed 
between the low and high states of 3C~279 (Maraschi et al. 1994). 
This effect is shown in Fig.2. Steep 
$\gamma$ ray spectra may thus be responsible for the non-detection at 
MeV/GeV energies of several blazars showing apparent superluminal motion. For 
NRAO~140, the X-ray flux and spectral index ($I_{1 keV}$=1.78$\times$ 
10$^{-3}$ photons / (cm$^2$ s keV) in the observer's frame, 
$\alpha_x$=1.73) reported by Turner et al. 
(1995) implies that the gamma-ray spectrum must be steeper than 
$\alpha_{\gamma}$=2.4 for a 
break energy of 10~MeV to give a flux below the EGRET detection threshold. 
For a break energy of 1~MeV, the non-detection requires $\alpha_{\gamma}$ 
$< 2.1$. 

However, six out of nine of the superluminal sources in the sample 
presented in von Montigny et al. (1995a) have EGRET spectra steeper than 
$\alpha_{\gamma}$=2. If we assume that the spectra of these sources 
do not change even during a $\gamma$-ray flare, this high detection 
rate indicates that a significant part of the EGRET sources either have a 
hard X-ray spectrum, or a spectral break occurs 
at energies higher than 1-10~MeV as derived by McNaron-Brown et al. (1995).

\section{Summary}

NRAO~140 has not been detected by any of the instruments on CGRO despite 
being considered a good candidate due to its X-ray brightness, hard 
X-ray spectrum and 
apparent superluminal motion. We argue that the source may have an 
X-ray/$\gamma$-ray spectrum similar to 3C~273, but is not detected since it 
is about 10 times dimmer than 3C~273. Accordingly, 
we conclude the non-detection of many superluminal sources may be 
due to their steep gamma-ray spectra and low break energies, not 
necessarily their hypothetical 
inactivity. 3C~34.47, which has similar properties as NRAO~140 
(Ohashi et al. 1992) and has also not yet been detected by EGRET 
(v. Montigny et al. 1995b), may be another typical example. 
High-sensitivity observations at hard X-rays with XTE and SAX can help 
to address this issue in the future.

\acknowledgements

MM acknowledges the support at MPE during the Comptel and EGRET data 
analysis, especially from W. Collmar and H.D. Radecke. We thank A. Kraus 
for discussion and data reduction of the radio data. This work was 
supported by DARA grant 50 OR 92054.

\end{document}